\documentclass[3p,times,procedia]{elsarticle}
\usepackage{nupha_ecrc}

\volume{00}

\firstpage{1}

\journalname{Nuclear Physics A}

\runauth{}

\jid{nupha}

\jnltitlelogo{Nuclear Physics A}

\usepackage{amssymb}

\usepackage[figuresright]{rotating}

\usepackage{epsfig}
\usepackage{color}
\usepackage{url}
\usepackage{amsmath}
\usepackage{multirow}
\def\bea#1\eea{\begin{align}#1\end{align}}

\newcommand{\bef}{\begin{figure}[htb]\centering}
\newcommand{\eef}{\end{figure}}

\begin{document}

\begin{frontmatter}




\title{Quenching of vector boson-tagged jets at the LHC}


\author[NU,ANL]{Hongxi Xing}

\address[NU]{Department of Physics and Astronomy, Northwestern University, Evanston, Illinois 60208, USA}
\address[ANL]{High Energy Physics Division, Argonne National Laboratory, Argonne, Illinois 60439, USA}


\begin{abstract}
Electroweak boson-tagged jet measurements provide a promising experimental channel to accurately study the physics of jet production and propagation in dense QCD medium. In this talk, we present theoretical predictions for the nuclear-induced attenuation of the differential cross section for isolated photon-tagged and $\rm Z^0$-tagged jet production in heavy ion collisions, and provide theoretical interpretations to the recent LHC data. We demonstrate quantitatively the significance of collisional and radiative energy losses, as revealed in the strong momentum asymmetry $d\sigma/dx_{VJ}$ and nuclear modification $\rm I_{AA}$ in central lead-lead reactions. 
\end{abstract}

\begin{keyword}
Jets \sep jet quenching 


\end{keyword}

\end{frontmatter}


\section{Introduction}
Vector boson-tagged jet production in heavy ion reactions is considered to be a ``golden channel'' for the study of jet quenching and the extraction of the properties of the hot dense medium. In this process, the tagging vector boson escapes the QCD medium unscathed (without strong interaction). It has been confirmed at this conference that the nuclear modification factors for $\gamma$ and $Z^0$ boson in Pb+Pb collisions at the LHC are consistent with unity. On the other hand, the parton shower that recoils opposite the tagging boson is quenched as it propagates through the hot dense medium. This channel provides very tight constraints on the energy and flavor origin of the away-side prompt jets in heavy ion reactions, and enables us to quantify precisely the flavor dependence of parton energy loss effects and to constrain the properties of the hot dense medium.

Recently, approximately back-to-back isolated $\gamma$+jet and $Z^0$+jet production in Pb+Pb collisions at $\sqrt{s_{NN}}=5.02$ TeV have been measured at the LHC by the ATLAS and CMS collaborations~\cite{Sirunyan:2017jic,CMS:2016ynj,ATLAS:2016tor}. Motivated by these new measurements, we present in this talk our theoretical calculations and comparisons to the experimental data~\cite{Kang:2017xnc}. In our analysis, we include both collisional \cite{Neufeld:2011yh, Neufeld:2014yaa} and radiative \cite{Gyulassy:2000fs} energy loss effects. In particular, we evaluate in both p+p and Pb+Pb collisions the transverse momentum imbalance ${\rm x_{JV}}$ distribution, where ${\rm x_{JV}} = p_T^J/p_T^V$ with $p_T^J$ and $p_T^V$ the transverse momentum of the recoiling jet and the tagging vector boson, respectively. In the analysis of ${\rm x_{JV}}$ distribution, we show the sensitivity of this observable to the strength of jet-medium interactions. We also calculate the nuclear modification factor ${\rm I_{AA}}$ for $\gamma$+jet and compare to the CMS measurements, through which we show the relative contributions of radiative and collisional energy losses of typical energy jets.  
\section{Quenching of photon-tagged and $Z^0$-tagged jet in heavy ion reactions at the LHC}
\label{pp}
We use Pythia 8~\cite{Sjostrand:2007gs} to evaluate the isolated photon-tagged and $Z^0$-tagged jet differential cross section in p+p collisions, where the vector boson (isolated-photon and $Z^0$-boson) and jet are selected according to the desired kinematics to match the experimental measurements. To validate Pythia simulations, we checked against CMS measurements for both photon-tagged and $Z^0$-tagged jets in p+p collisions at the LHC at $\sqrt{s}=7$~TeV.
 
In order to implement the energy loss effects through the medium-induced parton shower on vector boson-tagged jet production in Pb+Pb collisions at the LHC, we need the detailed information on the flavor origin of $Z^0$+jet and $\gamma$+jet production in p+p collisions. In Fig.~\ref{fig:pp-frac}, we show the fractional contribution of the two subprocesses in $Z^0$+jet (left) and $\gamma$+jet (right) production in p+p collisions. We find that the contribution from Compton scattering ($q(\bar q)+g\to V+q$) is dominant (around 80\%) for a wide $p_T$ range. This implies that in heavy ion collisions at LHC energies, the medium modification of V+jet production is dominated by quark energy loss, therefore, it serves as a golden channel to constrain the quark energy loss mechanism in hot dense medium. 
\bef
\psfig{file=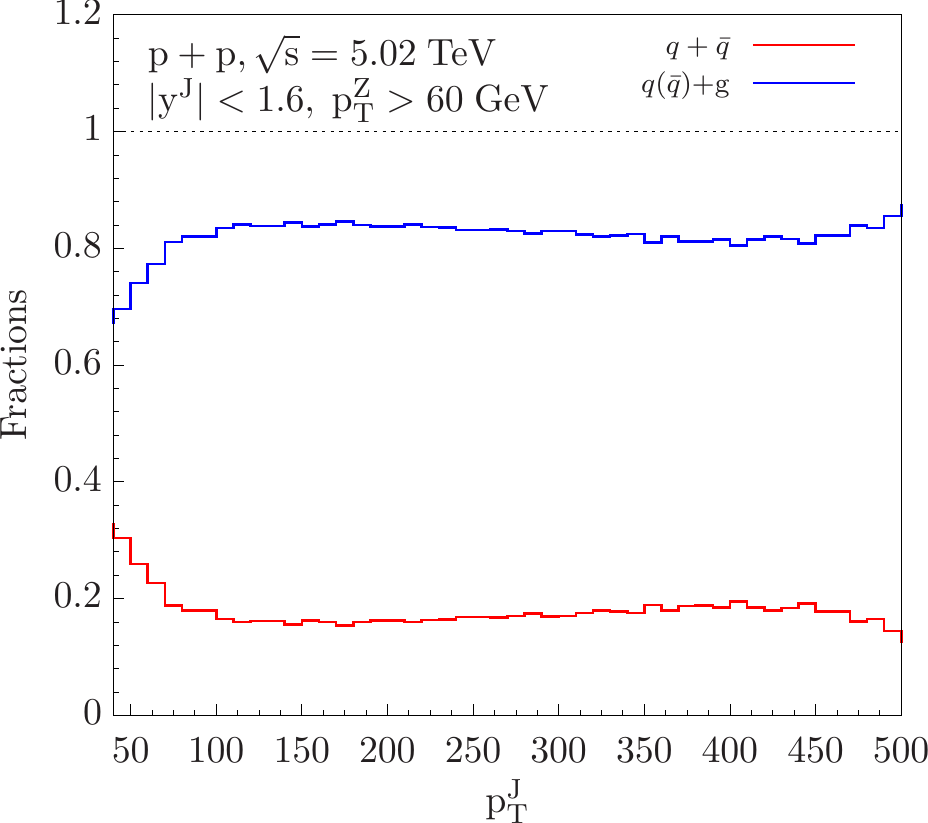, width=2.5in}
\hskip 0.5in
\psfig{file=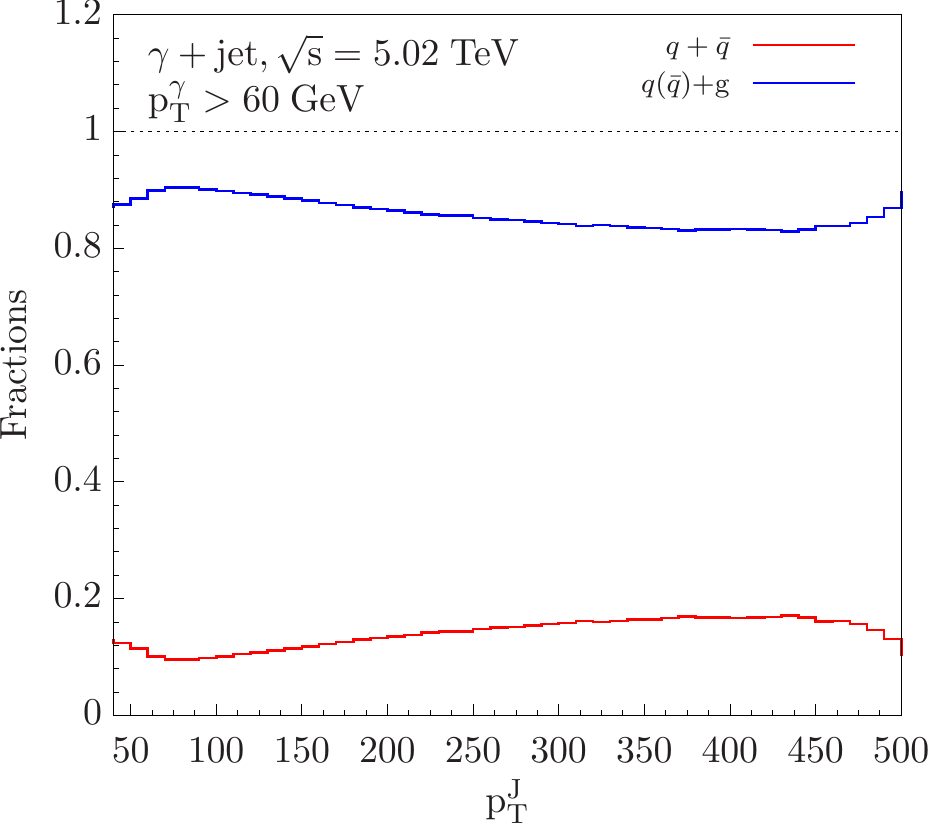, width=2.5in}
\caption{The fractional contributions of different subprocesses to the $Z^0$+jet (left) and isolated-$\gamma$+jet (right) production cross sections in p+p collisions at $\sqrt{s}=5.02$ TeV. } 
\label{fig:pp-frac}
\eef

We evaluate the medium modified vector boson-tagged jet cross section in nucleus-nucleus collisions in the soft gluon emission radiative energy loss approximation, which reads
\begin{eqnarray}
\frac{1}{\langle N_{bin}\rangle}\frac{d\sigma^{AA}}{dp_{T}^{V}dp_{T}^{J}} 
= \sum_{q,g} \int_0^1 d\epsilon \frac{P_{q,g}(\epsilon)}{1-f_{q,g}^{\rm loss}(R)\, \epsilon}
\frac{ d\sigma^{NN}_{q,g}  \left[ p_{T}^{V}, p_{T}^{J}/ \left(1-f_{q,g}^{\rm loss}(R)\,  \epsilon \right) \right] }
{dp_{T}^{J}dp_{T}^{V}} \, ,
\label{eq:modify}
\end{eqnarray}
where $P_{q,g}(\epsilon) $ is the probability distribution that a fraction $\epsilon$ of the energy of the parent parton is redistributed through medium-induced 
bremsstrahlung, it can be calculated from the medium-induced gluon radiative spectrum $\frac{dN^g_{q,g}(\omega,r)}{d\omega dr}$ of the parent quarks and gluons \cite{Vitev:2005he}.
To take into account how much of the energy of the medium-induced parton shower actually lost outside of the jet cone, we have included in Eq. (\ref{eq:modify}) the factor $f_{q,g}^{\rm loss}(R)$ \cite {Vitev:2008rz} in the evaluation of the differential cross section. 
 
From the evalulated 3-D $p_T$ distribution of the differential cross section, one can calculate the transverse momentum imbalance $\rm x_{JV}$ distribution as follows 
\bea
\frac{d\sigma}{d{\rm x_{JV}}} =& \int_{p_{T}^{J, \rm min}}^{p_{T}^{J, \rm max}} d p_{T}^J
\frac{p_T^J}{\rm{x}_{JV}^2}   \frac{d\sigma(p_T^V=p_T^J/{\rm x_{JV}}, p_T^J)}{dp_T^V dp_T^J} \;,
\label{sigAA}
\eea
where $p_{T}^{J, \rm min}$ and $p_{T}^{J, \rm max}$ are matched to the desired cuts in the  experimental measurements. 

In Fig. \ref{fig:XJZ} we show the normalized momentum imbalance distributions for the $Z^0$+jet production in both p+p and Pb+Pb collisions, and compare to the CMS measurements at the LHC~\cite{Sirunyan:2017jic}. The two plots show the unsmeared (left) and smeared (right) results, respectively. One can see that the $\rm x_{JZ}$ distribution from unsmeared Pythia 8 simulation is narrower than the one measured by the CMS experiment for the p+p reference, while the smearing leads to better agreement with the data.     
The results of our theoretical calculations in Pb+Pb collisions are shown for two different jet-medium coupling strengths $g=2.0$ (green) and $g=2.2$ (magenta). One can clearly see the downshift of ${\rm x_{JV}}$ as compared to p+p collisions and the sensitivity to the jet-medium coupling strength. The downshift of ${\rm x_{JV}}$ can be explained by the redistribution of the energy of away-side parton shower, while the $Z^0$-boson escapes out of the medium without strong interactions. The energy redistribution on the recoil side reduces the jet transverse momentum and results in the downshift of the ${\rm x_{JV}}$ distribution in Pb+Pb collisions, as observed in experimental data.
\bef
\psfig{file=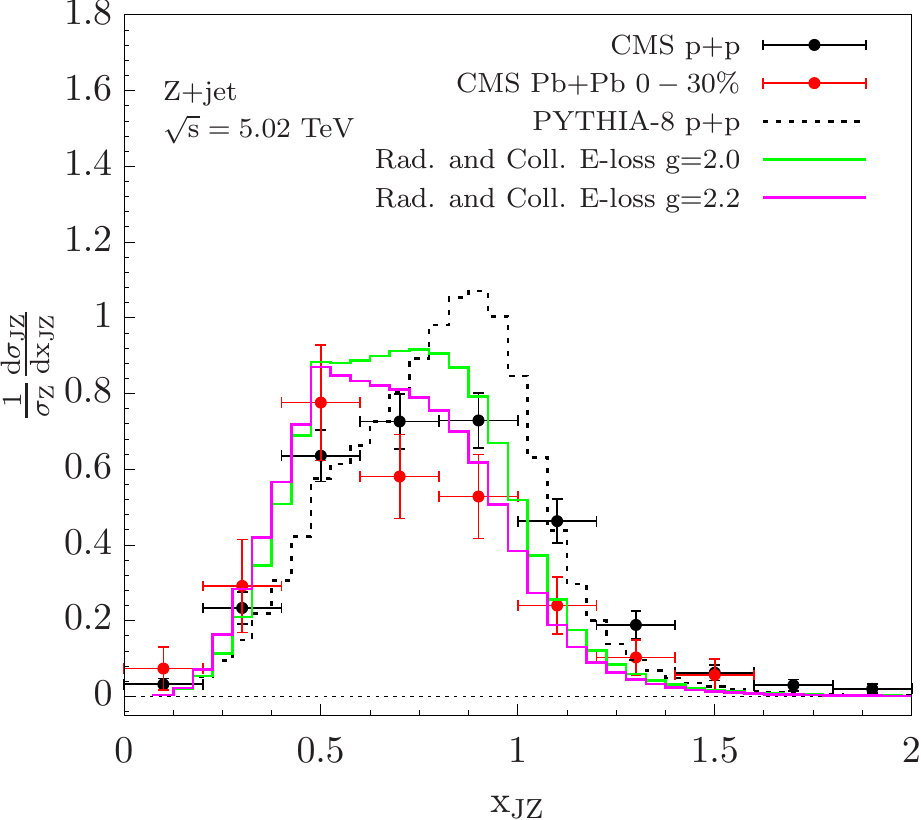, width=2.5in}
\hskip 0.5in
\psfig{file=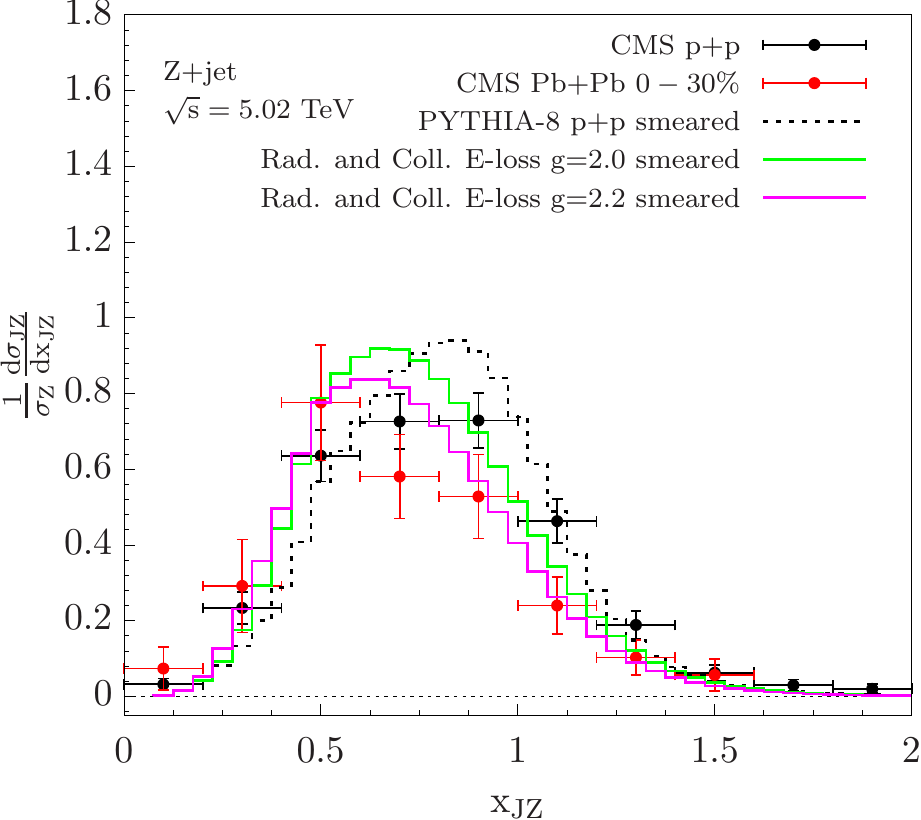, width=2.5in}
\caption{The $Z^0$-tagged jet asymmetry distribution at $\sqrt{s}=5.02$ TeV in p+p (black) and Pb+Pb (red) collisions at the LHC. The histograms shown in the left and right plots correspond to the unsmeared and smeared results, respectively.}
\label{fig:XJZ}
\eef

Isolated-$\gamma$-tagged jet production in p+p and Pb+Pb collisions are also evaluated and compared to CMS and ATLAS measurements as shown in Figs.~\ref{fig:xjphoton_cms}. Similar to what we observed in $Z^0$+jet production, the theoretical results of the difference between $ {\rm x_{J\gamma}}$ distributions in p+p and Pb+Pb collisions are quite compatible with what is seen in experimental data. Again, the difference can be explained by the nature of parton energy loss effects.  
\bef
\psfig{file=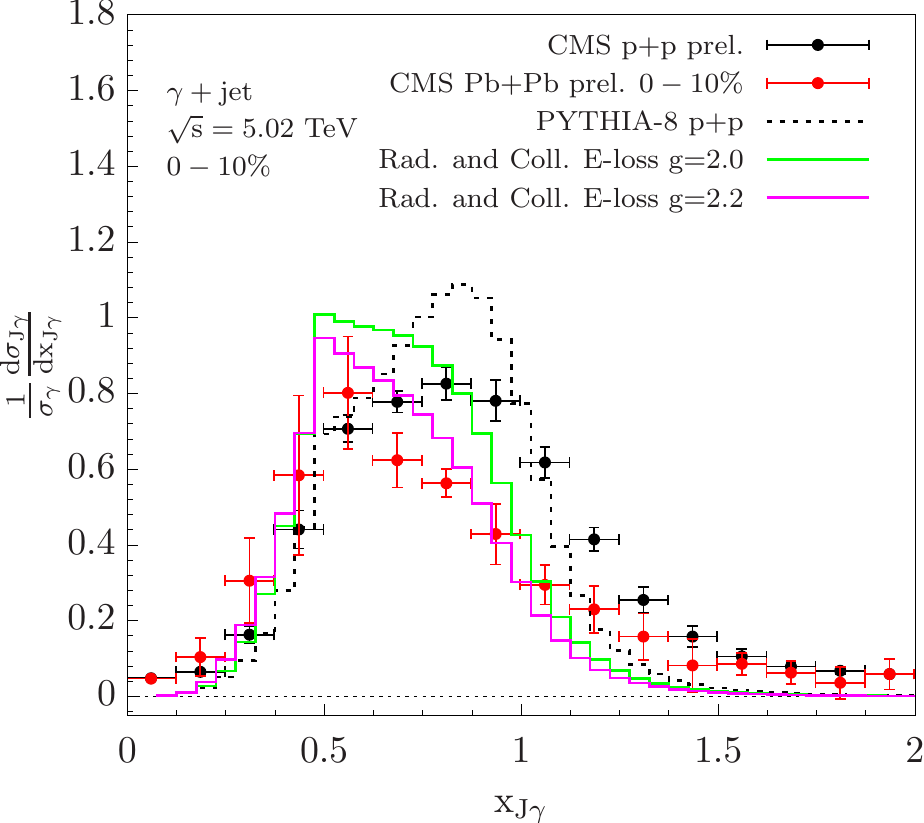, width=2.5in}
\hskip 0.5in
\psfig{file=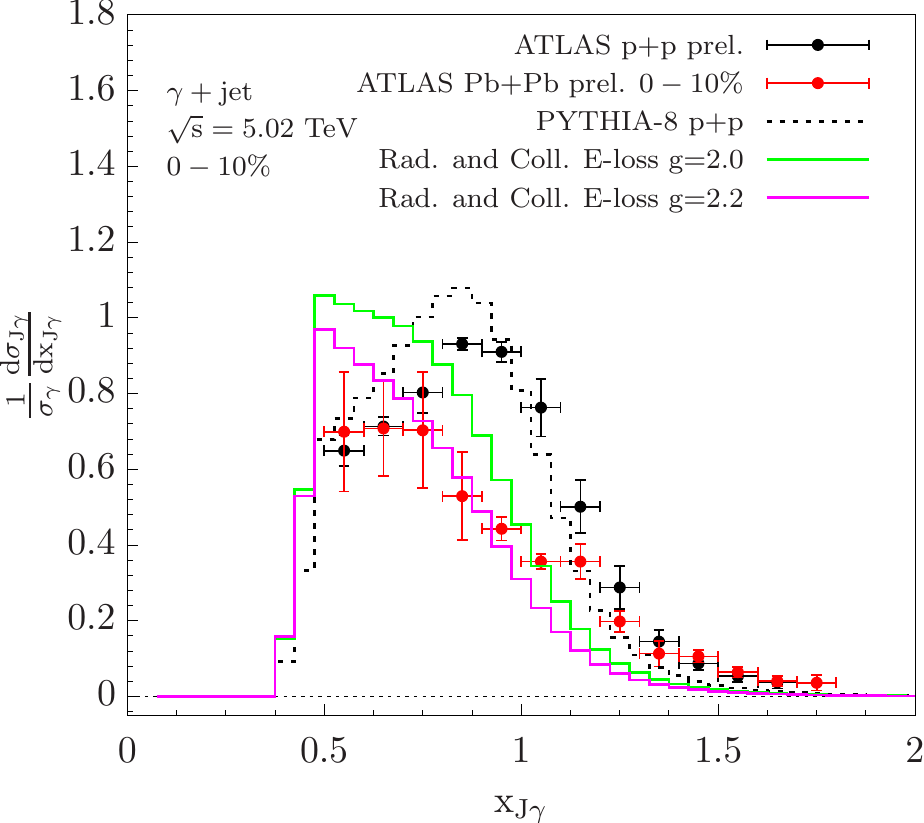, width=2.5in}
\caption{The isolated photon-tagged jet asymmetry distributions are shown and compared to CMS (left) and ATLAS (right) measurements~\cite{CMS:2016ynj, ATLAS:2016tor}.}
\label{fig:xjphoton_cms}
\eef

To further quantify the nuclear modification effects in isolated $\gamma$+jet production, we show in Fig. \ref{fig:IAA-photonjet} the theoretical results of the nuclear modification factor ${\rm I_{AA}}$ and the comparison to CMS experimental data \cite{Sirunyan:2017jic}. The energy loss effects are shown differentially in a combination of two different jet-medium coupling strengths, $g=2.0$ and $g=2.2$, as well as with and without the collisional energy loss effects. As one can see in Fig.~\ref{fig:IAA-photonjet}, there is a sensitive kinematical dependence of ${\rm I_{AA}}$. The strongest suppression is observed along the diagonal region of the transverse momenta of the trigger $\gamma$ and the recoil jet: $p_T^{\gamma} \approx p_T^{J}$, which can be naturally explained by the steeper falling cross section in the transverse momenta diagonal region, the suppression in the region $p_T^J>p_T^{\gamma}$ and enhancement in $p_T^J<p_T^{\gamma}$ is characteristic of in-medium tagged-jet dynamics. 
\bef
\psfig{file=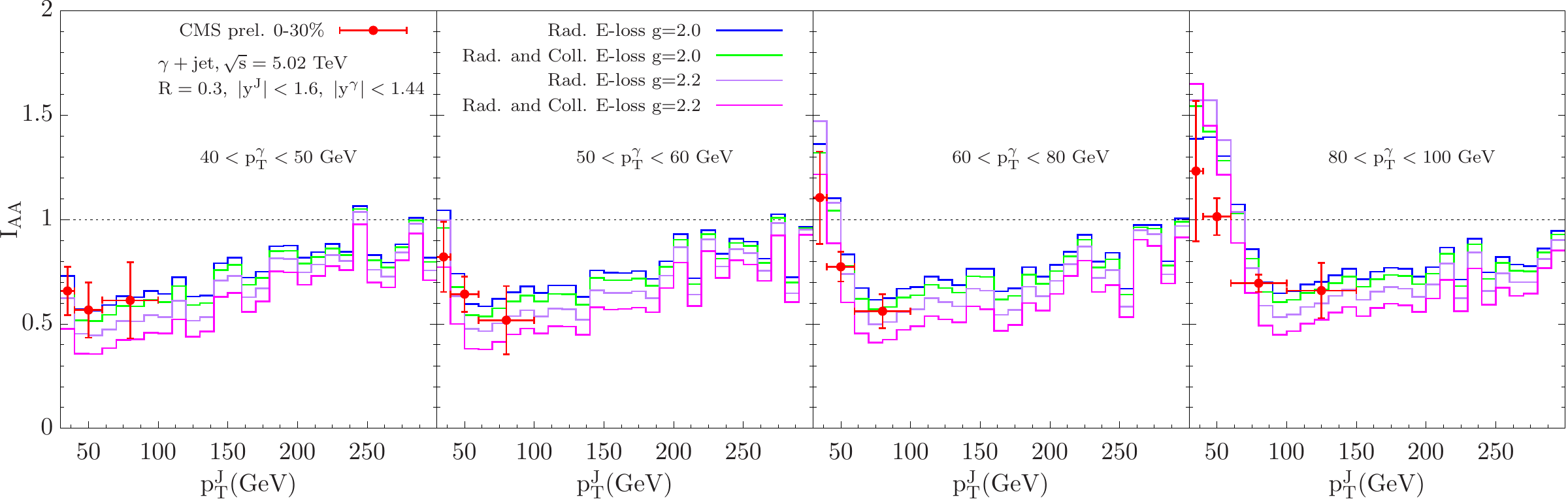, width=5.9in}
\caption{The transverse momentum cuts dependence of ${\rm I_{AA}}$ is shown in comparison to CMS data. We have chosen four different setups of energy loss effects: with and without collisional energy loss for both $g=2.0$ and $g=2.2$.}
\label{fig:IAA-photonjet}
\eef


\section{Conclusions}
\label{summary}
In conclusion, we presented in this talk a timely study of vector boson-tagged (either isolated $\gamma$ or $Z^0$)  jet production in p+p and Pb+Pb collisions at the LHC. As can be seen from Pythia 8 simulations for p+p collisions, both $\gamma$-tagged and $Z^0$-tagged jets are very effective in selecting prompt quark-initiated jets and are, thus,  excellent channels for the study of flavor dependence of parton energy loss. In Pb+Pb collisions, we have evaluated the transverse momentum imbalance ${\rm x_{JV}}$ distribution modification and the tagged jet nuclear modification factor ${\rm I_{AA}}$ within the traditional energy loss approach (soft gluon radiation limit). We found reasonably good agreement between the theoretical results and experimental measurements of these observables.  This agreement supports our understanding of the in-medium parton shower formation and related QCD dynamics, as encoded in these calculations.   

I thank Z. Kang and I. Vitev for collaboration, and the MIT CMS group for providing us with a detector resolution smearing function. This research is supported by the U.S. Department of Energy under Contract Nos. DE-FG02-91ER40684 and DE-AC02-06CH11357. 





\bibliographystyle{elsarticle-num}
\bibliography{<your-bib-database>}

\begin{thebibliography}{10}

\bibitem{Sirunyan:2017jic}
CMS, A.~M. Sirunyan {\em et~al.},
\newblock (2017), arXiv:1702.01060.

\bibitem{CMS:2016ynj}
CMS, C.~Collaboration,
\newblock (2016).

\bibitem{ATLAS:2016tor}
ATLAS, T.~A. collaboration,
\newblock (2016).

\bibitem{Kang:2017xnc} 
  Z.~B.~Kang, I.~Vitev and H.~Xing,
  arXiv:1702.07276 [hep-ph].

 \bibitem{Neufeld:2011yh}
R.~B. Neufeld and I.~Vitev,
\newblock Phys. Rev. {\bf C86}, 024905 (2012), arXiv:1105.2067.
  
 \bibitem{Neufeld:2014yaa}
R.~B. Neufeld, I.~Vitev, and H.~Xing,
\newblock Phys. Rev. {\bf D89}, 096003 (2014), arXiv:1401.5101.

\bibitem{Gyulassy:2000fs}
M.~Gyulassy, P.~Levai, and I.~Vitev,
\newblock Phys.Rev.Lett. {\bf 85}, 5535 (2000), arXiv:nucl-th/0005032.

\bibitem{Sjostrand:2007gs}
T.~Sjostrand, S.~Mrenna, and P.~Z. Skands,
\newblock Comput. Phys. Commun. {\bf 178}, 852 (2008), arXiv:0710.3820.
  
\bibitem{Vitev:2008rz}
I.~Vitev, S.~Wicks, and B.-W. Zhang,
\newblock JHEP {\bf 11}, 093 (2008), arXiv:0810.2807.
  
\bibitem{Vitev:2005he} 
  I.~Vitev,
  Phys.\ Lett.\ B {\bf 639}, 38 (2006)
  doi:10.1016/j.physletb.2006.05.083
  [hep-ph/0603010].
  
  
\end{thebibliography}



\end{document}